\documentclass[12pt]{iopart}
\usepackage{graphicx}

\begin{document}

\title[ ]
{Model construction and superconductivity analysis 
of organic conductors $\beta$-(BDA-TTP)$_{2}M$F$_{6}$ ($M$=P, As, Sb, Ta) 
based on first principles band calculation}

\author{H Aizawa$^1$\footnote{Author to whom any correspondence should be addressed}, 
K Kuroki$^2$, S Yasuzuka$^3$ and J Yamada$^4$}

\address{$^1$ Institute of Physics, Kanagawa University, 
 Yokohama, Kanagawa 221-8686, Japan}
\address{$^2$ Department of Engineering Science, The University of Electro-Communications, 
 Chofu, Tokyo 182-8585, Japan}
\address{$^3$ Faculty of Engineering, Hiroshima Institute of Technology, 
 Hiroshima 731-5193, Japan}
\address{$^4$ Department of Material Science, University of Hyogo, 
 Ako-gun, Hyogo 678-1297, Japan}

\ead{aizawa@kanagawa-u.ac.jp}

\begin{abstract}
We perform a first principles band calculation for a group of quasi-two-dimensional 
organic conductors $\beta$-(BDA-TTP)$_{2}M$F$_{6}$ ($M$=P, As, Sb, Ta). 
The \textit{ab-initio} calculation shows that 
the density of states (DOS) is correlated with 
the band width of singly occupied (highest) molecular orbital (SOMO), 
while it is not necessarily correlated with the unit cell volume.
The direction of the major axis of the cross section of the Fermi surface 
lies in the $\Gamma$-B direction, 
which differs from that obtained by 
the extended H$\ddot{\rm u}$ckel calculation. 
Then, we construct a tight-binding model which accurately reproduces 
the \textit{ab-initio} band structure. 
The obtained transfer energies give 
smaller dimerization than in the extended H$\ddot{\rm u}$ckel band. 
As for the difference of the anisotropy of the Fermi surface, 
the transfer energies along the inter-stacking direction 
are smaller than those obtained in the extended H$\ddot{\rm u}$ckel calculation.
Assuming spin-fluctuation-mediated superconductivity, 
we apply random phase approximation (RPA) to a two-band Hubbard model. 
This two-band Hubbard model is composed of the tight-binding model 
derived from the first principles band structure and 
an on-site (intra-molecule) repulsive interaction taken as a variable parameter. 
The obtained superconducting gap 
changes sign four times along the Fermi surface like in a $d$-wave gap, 
and the nodal direction is different from that obtained 
in the extended H$\ddot{\rm u}$ckel model. 
Anion dependence of $T_{\rm c}$ 
is qualitatively consistent with the experimental observation.
\end{abstract}

\pacs{74.70.Kn, 71.15.Mb, 71.10.Fd, 74.20.-z}

\submitto{\NJP}

\maketitle

\section{Introduction} 
\label{Introduction}

\begin{figure}[!htb]
 \centering
 \includegraphics[width=14.0cm]{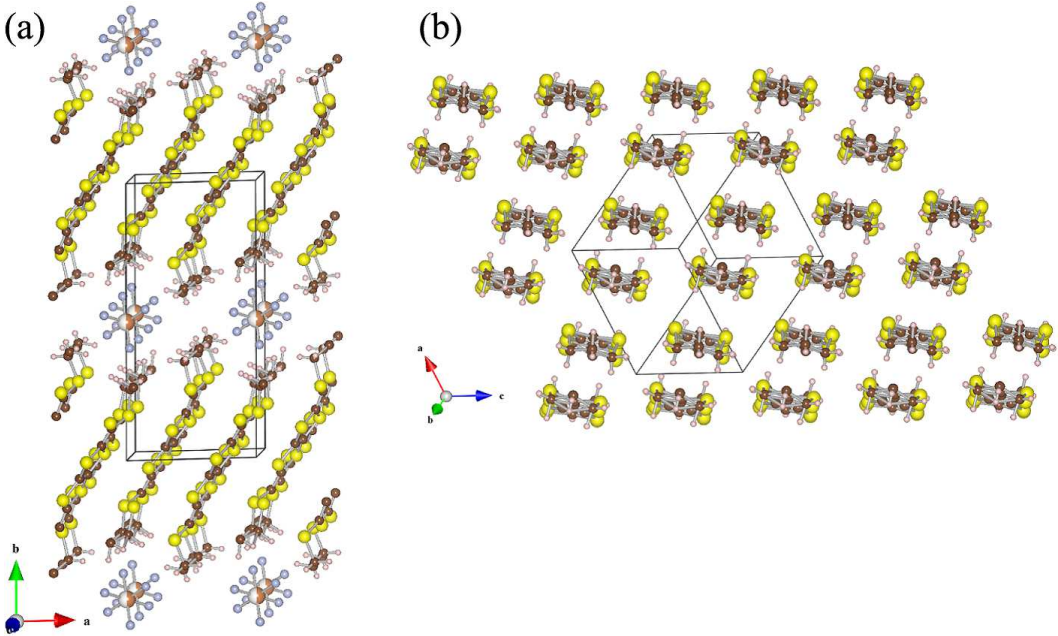}
 \caption{  
 Lattice structure of $\beta$-(BDA-TTP)$_{2}$SbF$_{6}$: 
 (a) the side view and (b) the top view of the conducting BDA-TTP layer. 
 }
 \label{fig1}
\end{figure}

Ever since their discovery, physics and chemistry of 
organic superconductors have attracted much attention 
\cite{Ardavan-Brown-JPSJ-81-011004}. 
One of the interesting features of these superconductors is the possibility of 
unconventional pairing arising from electron correlation effects. 
From this point of view, there have been efforts to control the 
strength of the electron correlation via chemical modification of molecules, 
and this has lead to the discovery of several new superconductors 
\cite{Yamada-Watanabe-JACS-104-5057}. 
Among those superconductors, here we focus on a group of organic conductors 
$\beta$-(BDA-TTP)$_{2}M$F$_{6}$ ($M=$P, As, Sb, Ta), 
where BDA-TTP is an abbreviation 
of 2,5-bis(1,3-dithian-2-ylidene)-1,3,4,6-tetrathiapentalene. 
BDA-TTP molecule is derived from BDH-TTP by stereochemical modification, 
where BDH-TTP is an abbreviation 
of 2,5-bis(1,3-dithiolan-2-ylidene)-1,3,4,6-tetrathiapentalene. 
While BDH-TTP salts show stable metallic behavior suggesting  
weak electron correlation, 
the BDA-TTP salts exhibit stronger electron correlation, which is expected 
from the band width narrowing due to the chemical modification. 
In fact, $\beta$-(BDA-TTP)$_{2}M$F$_{6}$ 
exhibit superconductivity with $T_{\rm c}$=7.5, 5.8, 5.9K for $M=$Sb, As, P, 
respectively \cite{Yamada-Watanabe-JACS-123-4174}, 
while $M=$Ta does not superconduct \cite{Yamada-Fujimoto-SM-153-373}. 
The conducting layer is sandwiched by insulating layers of anion $M$F$_{6}$ 
as shown in figure \ref{fig1}(a), 
from which we may expect these salts to be quasi-two-dimensional. 
The alignment of the BDA-TTP donors in the conducting layer 
takes the $\beta$-type configuration shown in figure \ref{fig1}(b). 
Measurements of Shubnikov-de Haas (SdH) oscillation and 
angular-dependent magneto-resistance oscillation (AMRO) 
for $\beta$-(BDA-TTP)$_{2}$SbF$_{6}$ have shown 
a two-dimensional Fermi surface \cite{Choi-Jobilong-PRB-67-174511} , 
which is consistent with the tight-binding band structure 
\cite{Yamada-Watanabe-JACS-123-4174} 
obtained from the extended H$\ddot{\rm u}$ckel calculation 
\cite{Mori-Kobayashi-BCSJ-57-627}. 
However, a very recent AMRO measurement shows 
that the direction of the anisotropy of the Fermi surface differs 
from the previous AMRO measurement, 
and suggests the direction of the major axis of the cross section 
lies in the $\Gamma$-B direction of the Brillouin zone (see figure \ref{fig2}) 
\cite{Yasuzuka-Koga-JPSJ-short-notes}. 
It is worth mentioning that 
the effective cyclotron mass takes a 
large value of $m_{c}^{*}/m_0 = 12.4 \pm 1.1$, 
which indicates that these materials are strongly correlated electron 
systems,  although the normal state exhibits a metallic nature. 

Measurements of the upper critical field ($H_{c2}$) of 
$\beta$-(BDA-TTP)$_{2}$SbF$_{6}$ 
have shown spin-singlet pairing 
\cite{Shimojo-Ishiguro-JPSJ-71-717,Tanatar-Ishiguro-PRB-71-024531}. 
The temperature dependence of the electronic specific heat has indicated 
presence of a nodal structure in the superconducting order parameter 
\cite{Shimojo-Ishiguro-JPSJ-71-717}, 
and a scanning tunneling microscope (STM) experiment also suggests 
presence of anisotropy in the gap, where the nodes (or the gap minima) 
lie in the $k_{a} \pm k_{c}$ direction 
\cite{Nomura-Muraoka-PhysicaB-404-562}. 
On the other hand, 
a recent study of the in-plane anisotropy of $H_{\rm c2}$ shows 
that the $H_{\rm c2}$ maxima  
appears when the magnetic field is applied 
in the $k_{a} \pm k_{c}$ direction, i.e., the 
nodal direction indicated in the STM measurement 
\cite{Yasuzuka-Koga-ICSM2010-5Ax-10}. 
This is puzzling because the $H_{\rm c2}$ maxima  usually 
appears when the field is applied in the antinodal direction.
A uniaxial pressure experiment  for $M=$As and Sb shows 
that $T_{\rm c}$ once increases and takes a maximum 
under the compression parallel to the stacking direction 
\cite{Ito-Ishihara-PRB-78-172506}. 

Theoretically, several studies have adopted the models derived from 
extended H$\ddot{\rm u}$ckel calculation, 
and analysis on superconductivity has also been performed 
assuming spin fluctuation mediated pairing 
\cite{Berk-Schrieffer-PRL-17-433,Moriya-Takahashi-Ueda-JPSJ-59-2905}. 
Nonoyama {\it et al.} 
\cite{Nonoyama-Maekawa-JPSJ-77-094703,comment-Nonoyama-et-al}
applied RPA to the two band model for $M=$As and Sb, 
while Suzuki {\it et al.} 
\cite{Suzuki-Onari-JPSJ-80-094704} 
applied the fluctuation exchange (FLEX) approximation 
to the original two-band model and 
the single-band dimer model for $M=$As and Sb, 
and discussed the pressure dependence. 
There, the $d$-wave like gap function has been obtained, 
whose nodal direction is consistent with the STM measurement 
\cite{Nomura-Muraoka-PhysicaB-404-562}, 
but not with the recent study of in-plane anisotropy of $H_{\rm c2}$ 
\cite{Yasuzuka-Koga-ICSM2010-5Ax-10}.

Thus, there are several controversies regarding the relation 
between experiments and theoretical studies based on the extended 
H$\ddot{\rm u}$ckel model. 
In the present paper, we study theoretically the organic conductors 
$\beta$-(BDA-TTP)$_{2}M$F$_{6}$ ($M=$P, As, Sb, Ta) based on a model 
constructed from {\it ab-initio} band calculation 
based on the density functional theory (DFT) 
exploiting the WIEN2K package \cite{WIEN2K}. 
In fact, the importance of model construction of molecular solids 
based on {\it ab-initio} calculation has recently pointed out 
in several studies 
\cite{Kandpal-Opahle-PRL-103-067004,Jeschke-Souza-PRB-85-035125,
Nakamura-Yoshimoto-JPSJ-78-083710}.
We point out the difference in the band structure and the Fermi surface 
between the {\it ab-initio} (non-empirical method) 
and the extended H$\ddot{\rm u}$ckel calculations (semi-empirical method), 
and its consequence to the superconducting gap mediated by spin fluctuations.
We also study the anion dependence of the band structure, 
and how it can affect the superconducting transition temperature.

\section{Method} 
\label{Method}

\subsection{{\it ab-initio} band calculation and model construction} 
\label{First principles band calculation} 

We perform an {\it ab-initio} band calculation 
using all-electron full potential linearized augmented plane-wave (LAPW) 
+ local orbitals (lo) method within WIEN2K \cite{WIEN2K}. 
This implements the DFT with different possible approximation 
for the exchange correlation potentials. 
The exchange correlation potential is calculated 
using the generalized gradient approximation (GGA). 

To attain convergence in the eigenvalue calculation, 
the single-particle wave functions in the interstitial region are 
expanded by plane waves with a cut-off of $R_{\rm MT} K_{\rm max}=3.0$,
where $R_{\rm MT}$ denotes the smallest muffin tin radius 
and $K_{\rm max}$ is the maximum value of $K$ vector 
in the plane wave expansion. 
For $\beta$-(BDA-TTP)$_{2}$SbF$_{6}$, for instance,
the muffin-tin radii are taken as 
1.74, 1.74, 1.62, 0.83, and 0.45 in  atomic units (au) 
for  Sb, F, S, C, and H, respectively. 
Thus $K_{\rm max}$ is 3/0.45$\approx$6.7, 
and the plane wave cutoff energy is 604.7 eV.
Calculations are performed using 7$\times$3$\times$9 $k$-points 
in the irreducible Brillouin zone. 
We adopt the lattice structure determined experimentally 
(including the coordinates of the H atoms)\cite{Yamada-Watanabe-JACS-123-4174}, 
and we do not relax the atomic positions in the calculation.

Having done the {\it ab-initio} calculation, we 
then construct a tight-binding model which accurately reproduces 
the \textit{ab-initio} band structure. 
From figure \ref{fig1}(b), 
it can be seen that the conducting BDA-TTP layer has 
two BDA-TTP molecules in a unit cell, so  
we regard one molecule as a site,  
and consider a two-band (two sites per unit cell) tight-binding model 
to fit the \textit{ab-initio} band structure. 
The tight-binding Hamiltonian we consider, $H_{0}$, is written in the form 
\begin{eqnarray}
 H_{0}=\sum_{\left< i \alpha: j \beta \right>, \sigma}
  \left\{ t_{i \alpha: j \beta} 
   c_{i \alpha \sigma}^{\dagger} c_{j \beta \sigma} + {\rm H. c.} 
  \right\}, 
\label{H0ij}
\end{eqnarray} 
where 
$i$ and $j$ are unit cell indices, 
$\alpha$ and $\beta$ specifies the sites in a unit cell, 
$c_{i \alpha \sigma}^{\dagger}$ ($c_{i \alpha \sigma}$ ) is 
a creation (annihilation) operator with spin $\sigma$ 
at site $\alpha$ in the $i$-th unit cell, 
$t_{i \alpha: j \beta}$ is the electron transfer energy 
between $(i, \alpha)$ site and $(j, \beta)$ site, and 
$\left< i \alpha: j \beta \right>$ represents 
the summation over the bonds corresponding to the transfer. 

By Fourier transformation, equation (\ref{H0ij}) is rewritten as 
\begin{eqnarray}
 H_{0}=\sum_{\textbf{\textit k}, \sigma, \alpha, \beta}
  \varepsilon_{\alpha \beta}\left( \textbf{\textit{k}} \right)
   c_{\textbf{\textit k} \alpha \sigma}^{\dagger} 
   c_{\textbf{\textit k} \beta \sigma}, 
\end{eqnarray}
where $\varepsilon_{\alpha \beta}\left( \textbf{\textit{k}} \right)$ is 
the site-indexed kinetic energy represented in $\textbf{\textit{k}}$-space. 
The band dispersion is given 
by diagonalizing $\varepsilon_{\alpha \beta}\left( \textbf{\textit{k}} \right)$, 
\begin{eqnarray}
 \varepsilon_{\alpha \beta}\left( \textbf{\textit{k}} \right)
  =\sum_{\gamma}
  d_{\alpha \gamma}\left( \textbf{\textit{k}} \right)
  d_{\beta \gamma}^{*}\left( \textbf{\textit{k}} \right) 
  \xi_{\gamma}\left( \textbf{\textit{k}} \right), 
  \label{eps-ab}
\end{eqnarray}
where $\xi_{\gamma}\left( \textbf{\textit{k}} \right)$ is 
the dispersion of the $\gamma$ band measured 
from the chemical potential, 
and $d_{\alpha \gamma}\left( \textbf{\textit{k}} \right)$ 
is the unitary matrix that gives the transformation.

In order to study unconventional superconductivity, 
we employ the two-band Hubbard model 
obtained by adding the on-site (intra-molecule) repulsive interaction 
to the tight-binding model derived from the {\it ab-initio} band structure. 
The Hubbard Hamiltonian, $H$, is 
\begin{eqnarray}
 H=H_{0}
  +\sum_{i \alpha} U
  n_{i \alpha \uparrow} n_{i \alpha \downarrow}
  \label{Hij}
\end{eqnarray} 
where $U$ is the on-site interaction and $n_{i \alpha \sigma}$ 
is the number operator of the electron on $\alpha$-site in the $i$-th unit cell. 
The on-site $U$ in the present study is fixed at a certain value. 
Although this choice does not have a quantitative basis, 
taking other values of $U$ does not affect qualitatively 
the conclusion of the present paper regarding the superconducting gap symmetry 
and the anion dependence of $T_c$.
Note also that we ignore the off-site (inter-molecule) repulsive interactions.
This is based on the result obtained in a previous theoretical work for 
the extended H$\ddot{\rm u}$ckel model of the present materials
\cite{Nonoyama-Maekawa-JPSJ-77-094703}.
There it has been shown that the introduction of moderate inter-molecule interactions 
only slightly modifies the position of the nodes of the gap. 
Although the model adopted here is different, 
we believe that the effect of the inter-site interaction should be similar 
as far as the off-site interaction values are moderate.

\subsection{Random phase approximation and superconducting gap equation} 
\label{Random phase approximation and superconducting gap equation} 

We apply RPA to the two-band Hubbard model given by equation (\ref{Hij}) as follows. 
The bare susceptibility matrix in the site-representation is given by 
\begin{eqnarray}
 \chi^{0}_{\alpha \beta} \left(\textbf{\textit{q}} \right)
 &=&\frac{-1}{N} \sum_{ \textbf{\textit{p}} } \sum_{ \gamma, \gamma' }
  d_{\alpha \gamma}\left( \textbf{\textit{p}}+\textbf{\textit{q}} \right) 
  d_{\beta  \gamma}^{*}\left( \textbf{\textit{p}}+\textbf{\textit{q}} \right) 
  d_{\beta  \gamma'}\left( \textbf{\textit{p}} \right)  
  d_{\alpha \gamma'}^{*}\left( \textbf{\textit{p}} \right) 
  \nonumber \\
 & &\times
   \frac{ f\left[ \xi_{\gamma} \left( \textbf{\textit{p}}+\textbf{\textit{q}} \right) \right]
         -f\left[ \xi_{\gamma'}\left( \textbf{\textit{p}} \right) \right] }
        { \xi_{\gamma} \left( \textbf{\textit{p}}+\textbf{\textit{q}} \right)
         -\xi_{\gamma'}\left( \textbf{\textit{p}} \right) }, 
 \label{chi0}
\end{eqnarray}
where $N$ is the total number of unit cells, and 
$f\left( \xi \right)$ is the Fermi distribution function
(this is the multi-site version of the Lindhard function). 
The indices $\alpha\beta$ on the left hand side 
means ($\alpha$, $\beta$)-element of 
the bare susceptibility matrix $\hat{{X}}^{0}$. 
Within  RPA, the spin and charge susceptibilities are obtained as 
\begin{eqnarray}
 \hat{{X}}^{\rm sp}
  &=&\left( \hat{I}-\hat{{X}}^{0} \hat{U} \right)^{-1} \hat{{X}}^{0},
  \label{chisp} 
  \\
 \hat{{X}}^{\rm ch}
  &=&\left( \hat{I}+\hat{{X}}^{0} \hat{U} \right)^{-1} \hat{{X}}^{0}, 
  \label{chich}
\end{eqnarray}
respectively. 
$\hat{{X}}^{\rm sp}$, $\hat{{X}}^{\rm ch}$, $\hat{{X}}^{0}$
are the spin, charge, and bare susceptibility matrices, respectively, 
$\hat{U}$ is the on-site interaction matrix, and $\hat{I}$ is the unit matrix.  
In a two-band system, 
$\hat{U}$, $\hat{X}^{0}$, $\hat{X}^{\rm sp}$ and $\hat{X}^{\rm ch}$ 
all become 2$\times$2 matrices. 
The ``spin susceptibility'' (see figure \ref{fig7}(a)) 
in the present paper is obtained 
as the larger eigenvalue of the 2$\times$2 spin  susceptibility matrix. 

By using these susceptibility matrices,  
the pairing interaction $\hat{V}^{\rm singlet}$ for the spin-singlet state 
is given as
\begin{eqnarray}
 \hat{V}^{\rm singlet} 
  &=& \hat{U}
     +\frac{3}{2}\hat{U}\hat{{X}}^{\rm sp}\hat{U}
     -\frac{1}{2}\hat{U}\hat{{X}}^{\rm ch}\hat{U}. 
  \label{pairing-int}
\end{eqnarray}
Using the pairing interaction given by equation (\ref{pairing-int}), 
we solve the linearized gap equation to obtain 
the transition temperature $T_{\rm c}$ and the superconducting gap function 
for the spin-singlet pairing state. 
The linearized gap equation within the weak-coupling theory is given by 
\begin{eqnarray}
 \lambda \varphi_{\alpha \beta} \left( \textbf{\textit{k}} \right) 
  &=&
  \frac{-1}{N}\sum_{ \textbf{\textit{k}}' }\sum_{\alpha' \beta'}\sum_{\gamma \gamma'}
  V^{\rm singlet}_{\alpha \beta} \left( \textbf{\textit{k}}-\textbf{\textit{k}}' \right)
  \nonumber \\
  &\times&
  d_{\alpha  \gamma }    \left( \textbf{\textit{k}}' \right)
  d_{\alpha' \gamma }^{*}\left( \textbf{\textit{k}}' \right) 
  d_{\beta   \gamma'}    \left(-\textbf{\textit{k}}' \right)
  d_{\beta'  \gamma'}^{*}\left(-\textbf{\textit{k}}' \right)
  \nonumber \\
  &\times&
  \frac{ -f\left[ \xi_{\gamma } \left( \textbf{\textit{k}}' \right) \right]
         +f\left[-\xi_{\gamma'} \left(-\textbf{\textit{k}}' \right) \right]}
       { \xi_{\gamma }\left( \textbf{\textit{k}}' \right)
        +\xi_{\gamma'}\left(-\textbf{\textit{k}}' \right)}
	\varphi_{\alpha' \beta'}\left( \textbf{\textit{k}}' \right), 
  \label{gap-eq}
\end{eqnarray}
where $\lambda$ is the eigenvalue of the linearized gap equation, 
and $\varphi_{\alpha \beta}\left( \textbf{\textit{k}} \right)$ 
is the ($\alpha$, $\beta$)-element of the gap function matrix. 
The transition temperature $T_{\rm c}$ 
is the temperature where $\lambda$ reaches unity. 
In a two-band system, 
$\hat{V}^{\rm singlet}$ and $\hat{\varphi}$ become 2$\times$2 matrices. 

Although RPA is quantitatively insufficient for evaluating 
the absolute value of $T_{\rm c}$, 
we expect this approach to be valid for 
(i) studying the form of the superconducting gap function assuming 
spin-fluctuation-mediated pairing, and 
(ii) qualitatively comparing the tendency toward superconductivity (or $T_c$) 
between the different anions. 
We take 128$\times$128 $k$-point meshes in the RPA calculation 
throughout the entire study.

\section{Results} 
\label{Results}

\subsection{{\it ab-initio} band calculation} 
\label{First principles band calculation} 

\begin{figure}[!htb]
 \centering
 \includegraphics[width=10.0cm]{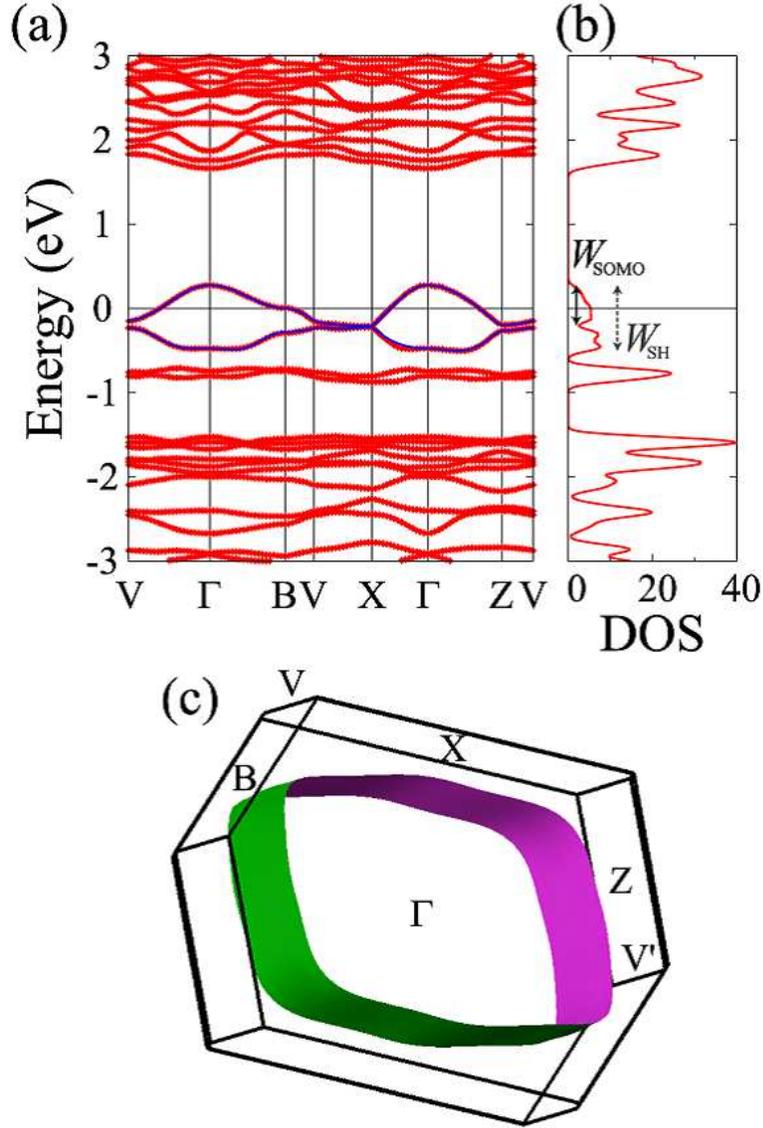}
 \caption{  
 (a) Calculated \textit{ab-initio} band structure, (b) the density of states, 
 and (c) the Fermi surface 
 of $\beta$-(BDA-TTP)$_{2}$SbF$_{6}$ at ambient pressure and room temperature, 
 where X(Z)-axis corresponds to the $a$($c$)-direction in the lattice structure. 
 In figure (a), the red curve represent the \textit{ab-initio} band dispersions 
 and the blue solid curves gives the tight-binding fit. 
 }
 \label{fig2}
\end{figure}

\begin{figure}[!htb]
 \centering
 \includegraphics[width=10.0cm]{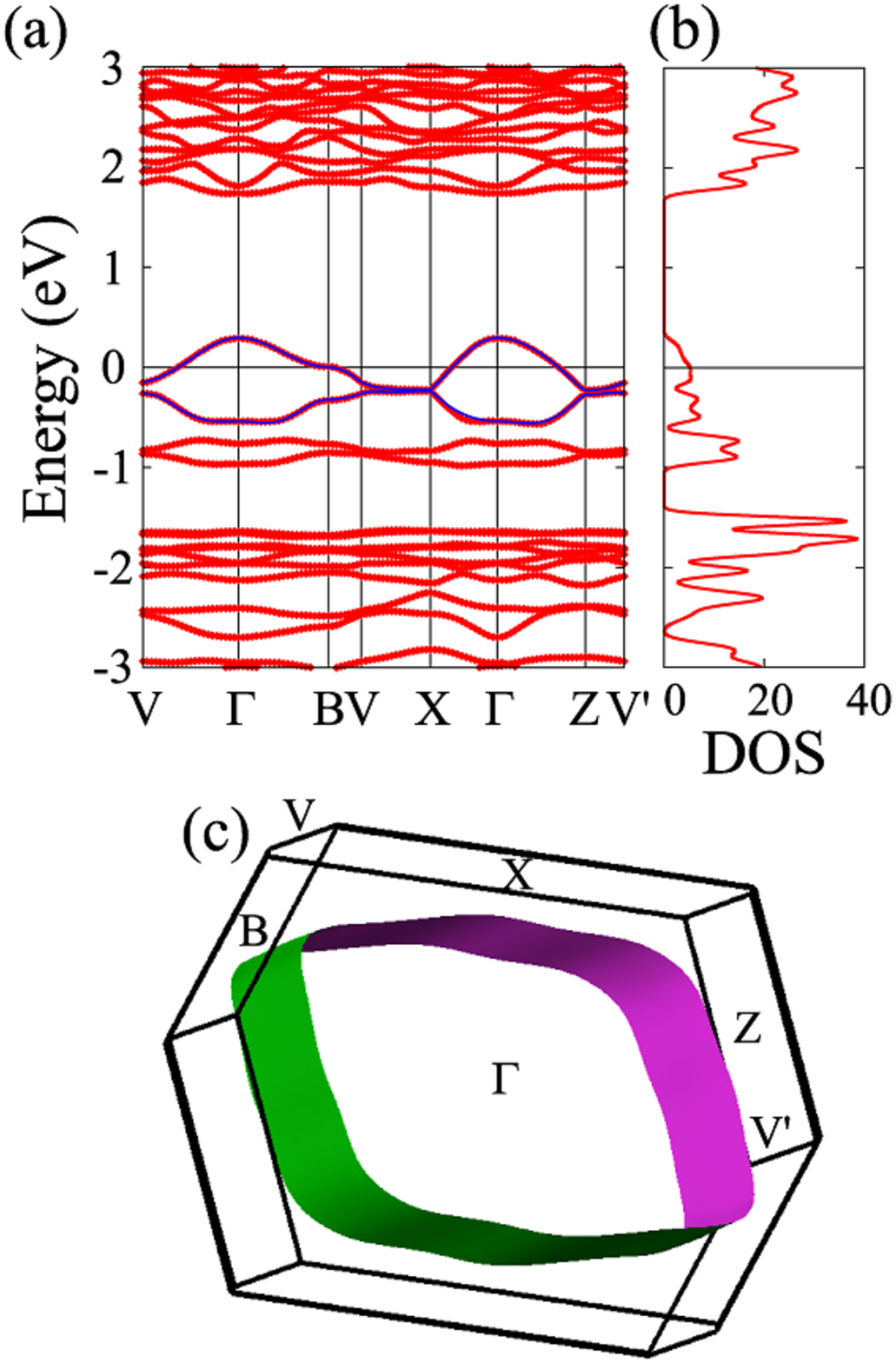}
 \caption{  
 Plots similar to Fig\ref{fig2} for 
$\beta$-(BDA-TTP)$_{2}$AsF$_{6}$ at ambient pressure and room temperature ; 
(a) Calculated \textit{ab-initio} band structure, (b) the density of states, 
and (c) the Fermi surface. 
 }
 \label{fig3}
\end{figure}

We show in figure \ref{fig2}(a) the \textit{ab-initio} band structure 
for $\beta$-(BDA-TTP)$_{2}$SbF$_{6}$ at ambient pressure and room temperature. 
It can be seen that the 
singly occupied (highest) molecular orbital (SOMO) and the 
highest-occupied molecular orbital (HOMO) 
near the Fermi level (zero energy) 
are isolated from other bands, 
although there are two bands somewhat close to the HOMO band, 
which correspond to HOMO$-$1 and HOMO$-$2. 
Density of states (DOS) in figure \ref{fig2}(b) also shows 
the isolation of the HOMO and SOMO from the other bands. 
From these results, 
it can be considered that the SOMO and HOMO bands are the target bands 
to construct a low energy tight-binding model. 
figure \ref{fig2}(c) shows the Fermi surface of the 
\textit{ab-initio} calculation, 
where the name of the 
$k$-points are presented only on the $k_{\rm Y} (k_{b})=0$ plane, 
e.g $\Gamma$=(0, 0, 0), Z=(0, 0, $\pi$/c ) and B=($\pi$/a, 0, $-\pi$/c ). 
We also show similar plots for $\beta$-(BDA-TTP)$_{2}$AsF$_{6}$ 
at ambient pressure and room temperature in figure \ref{fig3}. 
The band structure and the Fermi surface are similar between the 
two salts. $M=$P and Ta also give similar results (not shown).
The DOS also looks similar qualitatively, but its quantitative 
dependence on the anions will be discussed later.

The Fermi surface is clearly cylindrical, 
reflecting the strong two dimensionality of this salt. 
The direction of the anisotropy of the Fermi surface 
differs from that found in 
the previous extended H$\ddot{\rm u}$ckel calculation 
\cite{Yamada-Watanabe-JACS-123-4174,Yamada-Fujimoto-SM-153-373,
Nonoyama-Maekawa-JPSJ-77-094703,Suzuki-Onari-JPSJ-80-094704}. 
The origin of this difference is intimately related to 
including/excluding the 3d-orbital 
of the sulfur atoms of the TTP skeleton donors 
in the extended H$\ddot{\rm u}$ckel calculation 
\cite{Ouyang-Yakushi-JPSJ-67-3191,Kawamoto-Ashizawa-JPSJ-71-3059,
Drozdova-Yakushi-JSSC-168-497}. 
%
Actually, the shape of the Fermi surface 
given by the extended H$\ddot{\rm u}$ckel calculation in which 
the 3d-orbital of the sulfur atoms is excluded 
\cite{Kawamoto-private-comm-3d} 
is similar to the present \textit{ab-initio} result. 
Focusing on the comparison of the experiments 
with our results of the Fermi surface, 
the anisotropy of the Fermi surface differs from the previous experiment 
\cite{Choi-Jobilong-PRB-67-174511}, 
but a very recent AMRO measurement shows that 
the long-axis of the Fermi surface ellipse  
is in the $\Gamma$-B direction, 
which is at least qualitatively consistent with the present result 
\cite{Yasuzuka-Koga-JPSJ-short-notes}. 
However, the size of the Fermi surface is not consistent 
between this experiment and the present calculation, 
and this remains as a puzzle.

Now, one of the important issues we study in the present paper is the 
effect of substituting the anion. Here we consider the anion dependence of 
the band-width and the DOS at the Fermi level 
for $\beta$-(BDA-TTP)$_{2}M$F$_{6}$ ($M$=P, As, Sb, Ta) 
at ambient pressure and room temperature, 
and also for $\beta$-(BDA-TTP)$_{2}$SbF$_{6}$ at 12K. 
As for the band width, we introduce $W_{\rm SH}$ and $W_{\rm SOMO}$, 
where the former is the energy difference between the top of 
SOMO and the bottom of HOMO,  
and the latter is the band width of SOMO alone.
\begin{figure}[!htb]
 \centering
 \includegraphics[width=11.0cm]{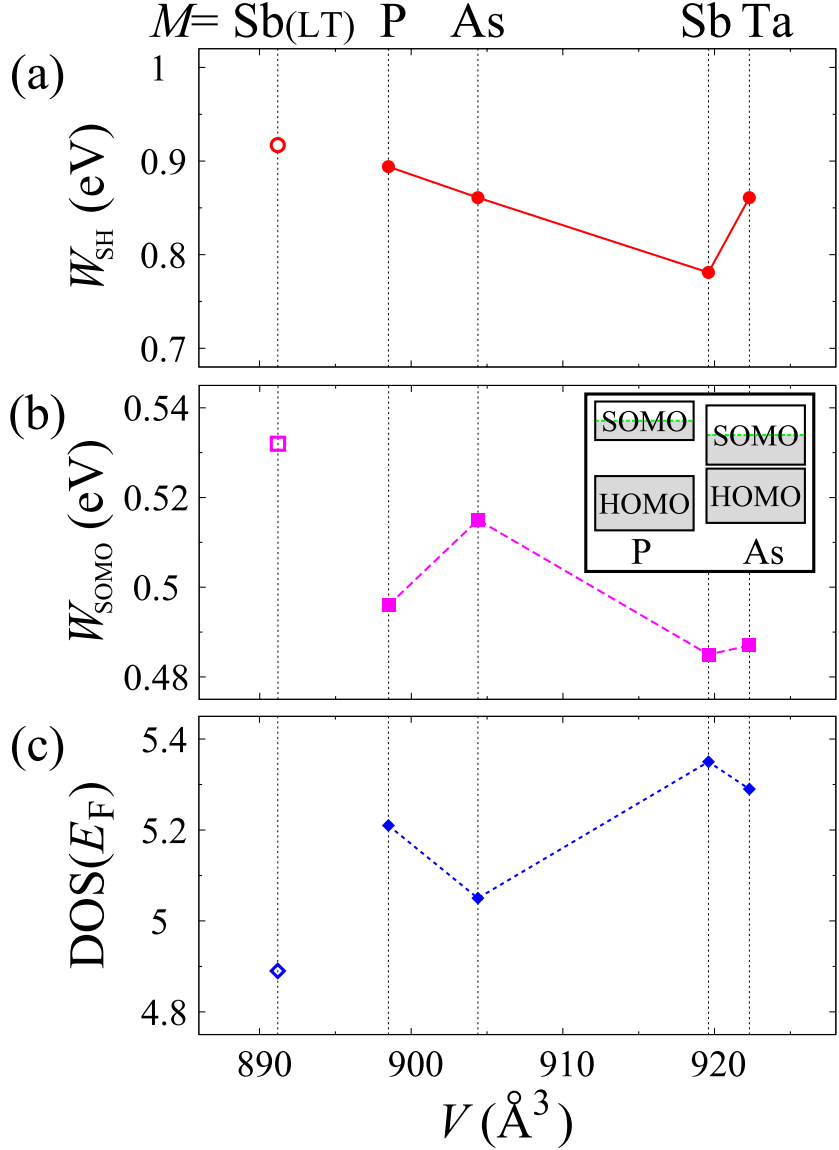}
 \caption{  
 (a) SOMO+HOMO band width, (b) SOMO band width, and (c) the density of states 
 against the unit-cell volume. 
 LT in parentheses for Sb means that the low temperature
lattice parameters are used. 
 The inset of (b) shows the schematic relation between SOMO and HOMO. 
 }
 \label{fig4}
\end{figure}
figure \ref{fig4}(a) shows that 
$W_{\rm SH}$ becomes smaller as the anion is enlarged from $M=$P to Sb,  
resulting in an increase of the unit-cell volume.
The trend can naturally be understood as the effect of the chemical pressure.
$M=$Ta, however, is an exception, where the increase of the unit-cell volume 
from $M=$Sb results in an increase in the band width. 
The relation between $M=$Sb and $M=$Sb(LT) shows that 
lowering the temperature results in an increases of $W_{\rm SH}$ 
due to thermal compression. 
We show in figure \ref{fig4}(b) $W_{\rm SOMO}$ against the unit-cell 
volume. 
The correlation between $W_{\rm SOMO}$ and the volume is not monotonic 
in contrast to the volume dependence of $W_{\rm SH}$,  
even within the pnictogens  $M=$P, As and Sb. 
The reason for this lies in the small dimerization gap 
between the SOMO and HOMO bands. 
Even though $W_{\rm SOMO}$ of $M=$As is larger than that of $M=$P, 
$W_{\rm SH}$ of As is smaller 
because the dimerization gap of As is smaller than that of P. 
A schematic figure is shown in the inset of figure \ref{fig4}(b), 
where the left (right) figure corresponds to $M=$P (As). 
We will come back to this point in the next subsection.
The volume dependence of the DOS at the Fermi level, DOS($E_{\rm F}$), 
varies in accord with $W_{\rm SOMO}$, not $W_{\rm SH}$, 
as seen in figure \ref{fig4}(c). 
It is natural that DOS is inversely related to $W_{\rm SOMO}$ 
because this directly measures the width of the band 
that intersects the Fermi level, 
but the point here is that $W_{\rm SOMO}$ is not directly 
correlated with $W_{\rm SH}$ and thus the unit cell volume.

\subsection{Effective tight-binding model}
\label{Effective model} 

\begin{figure}[!htb]
 \centering
 \includegraphics[width=11.0cm]{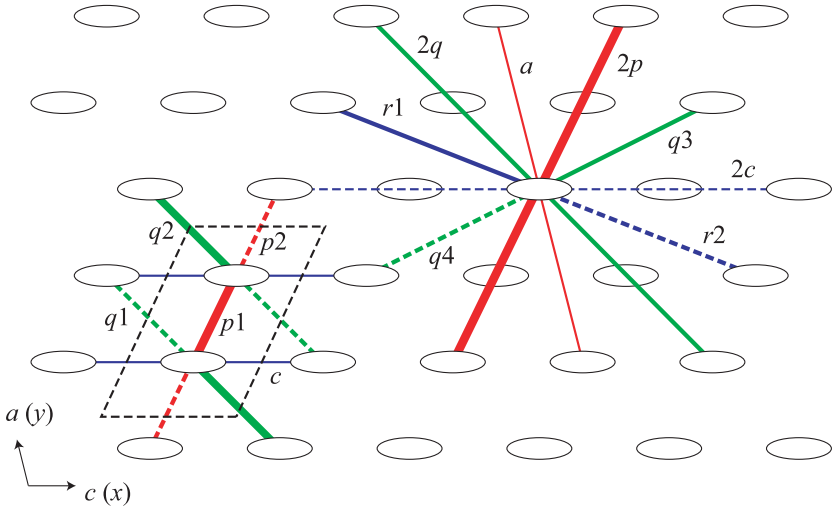}
 \caption{  
 The tight-binding model adopted to fit the \textit{ab-initio} band structure, 
 where an ellipse corresponds to a BDA-TTP molecule. 
 }
 \label{fig5}
\end{figure}

Figure \ref{fig5} shows the effective tight-binding model 
adopted to fit the \textit{ab-initio} band. 
In addition to the 
nearest-neighbor transfer energies (left part in figure \ref{fig5}, 
same notation is used as in previous studies), 
we need to introduce the next-nearest-neighbor transfer energies 
(right panel in figure \ref{fig5}) 
to reproduce the \textit{ab-initio} band more accurately.
Note that the stacking direction of the BDA-TTP molecules is 
taken in the $c+a$-direction of the lattice structure in our study, 
as shown in figure \ref{fig1}(b). 

\begin{figure}[!htb]
 \centering
 \includegraphics[width=12.0cm]{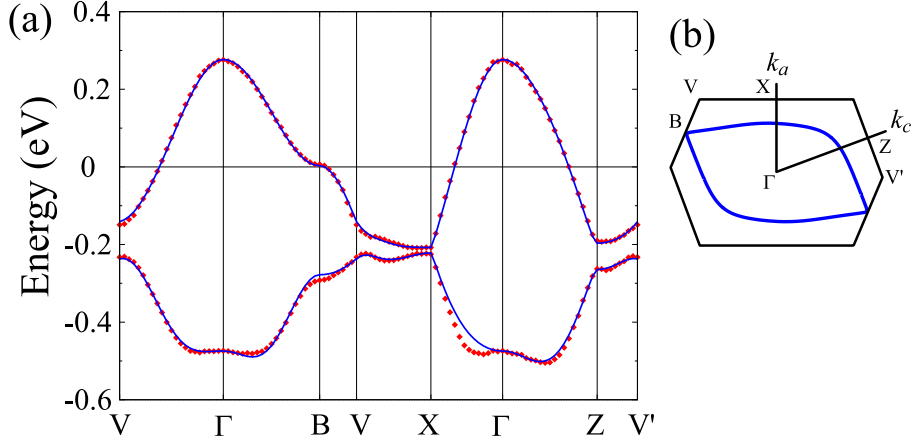}
 \caption{  
 Blue solid curves represent the band structure of the tight-binding model 
 which is fit to the \textit{ab-initio} band (red diamonds) 
 of $\beta$-(BDA-TTP)$_{2}$SbF$_{6}$ at room temperature. 
 }
 \label{fig6}
\end{figure}

We show  in figure \ref{fig6}(a) the band structure of the 
two band tight-binding model 
for $\beta$-(BDA-TTP)$_{2}$SbF$_{6}$ superposed to the 
\textit{ab-initio} band structure. 
It can be seen that the tight-binding model nicely reproduces the 
\textit{ab-initio} band structure. 
figure \ref{fig6}(b) shows the Fermi surface obtained from the model,  
which is essentially the same as the \textit{ab-initio} result  
in figure \ref{fig2}(c). 

Similar tight-binding fit is performed for all the anions, and 
the thereby determined transfer energies are summarized in Table \ref{tab1}. 
The bottom two lines of Table \ref{tab1} represent 
the magnitude of the dimerization, $t_{p1}/t_{p2}$, in the stacking direction, 
and the relative transfer 
perpendicular to that direction (i.e. the strength of the 
two dimensionality), $(t_{q1}+t_{q2})/(t_{p1}+t_{p2})$.
\begin{table}[!htb]
\centering
 \caption{List of the transfer energies in the unit of eV 
 for $\beta$-(BDA-TTP)$_{2}M$F$_{6}$, 
 where LT means a structural data measured in low temperature.} 
 \begin{tabular}{ l|c|c|c|c|c } \hline
  \hspace{1pt} $M$F$_{6}$ \hspace{1pt} & \hspace{0pt} SbF$_{6}$ (LT)  \hspace{0pt} &
  \hspace{5pt} PF$_{6}$   \hspace{5pt} & \hspace{5pt} AsF$_{6}$  \hspace{5pt}      & 
  \hspace{5pt} SbF$_{6}$  \hspace{5pt} & \hspace{5pt} TaF$_{6}$  \hspace{5pt}   \\ \hline \hline
  $t_{p1}$ (eV)    & -0.171   & -0.176   & -0.162   & -0.153   & -0.166   \\
  $t_{p2}$         & -0.155   & -0.141   & -0.149   & -0.126   & -0.138   \\
  $t_{q1}$         & -0.070   & -0.065   & -0.066   & -0.055   & -0.061   \\
  $t_{q2}$         & -0.082   & -0.082   & -0.071   & -0.071   & -0.083   \\
  $t_{c }$         &  0.009   &  0.003   &  0.008   &  0.007   &  0.004   \\
  $t_{2c}$         &  0.008   &  0.006   &  0.005   &  0.005   &  0.006   \\
  $t_{2p}$         &  0.017   &  0.015   &  0.018   &  0.021   &  0.015   \\
  $t_{a }$         & -0.0002  & -0.002   &  0.002   &  0.003   & -0.001   \\
  $t_{2q}$         &  0.007   &  0.006   &  0.007   &  0.005   &  0.006   \\
  $t_{q3}$         &  0.005   & -0.002   &  0.006   &  0.003   & -0.002   \\
  $t_{q4}$         &  0.005   &  0.006   &  0.003   &  0.006   &  0.009   \\
  $t_{r1}$         &  0.014   &  0.015   &  0.012   &  0.014   &  0.016   \\
  $t_{r2}$         &  0.015   &  0.008   &  0.011   &  0.008   &  0.008   \\ \hline \hline
  $t_{p1}/t_{p2}$  &  1.10    &  1.25    &  1.09    &  1.21    &  1.20    \\ 
  $\frac{t_{q1}+t_{q2}}{t_{p1}+t_{p2}}$  
                   &  0.466   &  0.464   &  0.441   &  0.452   &  0.474   \\ \hline
 \end{tabular}
 \label{tab1}
\end{table}
We can now quantitatively discuss the effect of the dimerization on 
the unit-cell volume dependence of $W_{\rm SOMO}$ and DOS($E_{\rm F}$). 
Table \ref{tab1} shows that 
$t_{p1}/t_{p2}$ obtained by using the structural data at room temperature 
is around 1.22 except for $M=$As, 
in which $t_{p1}/t_{p2}$ is 1.09. This means that the dimerization gap 
between the SOMO and HOMO bands is small in As, hence resulting in a
larger $W_{\rm SOMO}$ despite the smaller $W_{\rm SH}$.
This is consistent with the qualitative discussion given 
in subsection \ref{First principles band calculation}. 
From the viewpoint of the transfer energies, 
we consider that the difference of the anisotropy of the Fermi surface 
is due to the magnitude of the two-dimensionality, 
$(t_{q1}+t_{q2})/(t_{p1}+t_{p2})$. 
Our result shows that 
the average value of $(t_{q1}+t_{q2})/(t_{p1}+t_{p2})$ 
over the salts listed in Table \ref{tab1} is around 0.46, 
which is smaller than that obtained by 
the extended H$\ddot{\rm u}$ckel calculation 
\cite{Yamada-Watanabe-JACS-123-4174,Yamada-Fujimoto-SM-153-373,
Nonoyama-Maekawa-JPSJ-77-094703,Suzuki-Onari-JPSJ-80-094704}. 
Therefore, the present Fermi surface exhibits an anisotropy 
in which the $a+c$-direction is more conductive than in 
the extended H$\ddot{\rm u}$ckel result.

\subsection{Random phase approximation and the superconducting gap equation} 
\label{Random phase approximation and superconducting gap equation} 

\begin{figure}[!htb]
 \centering
 \includegraphics[width=15.5cm]{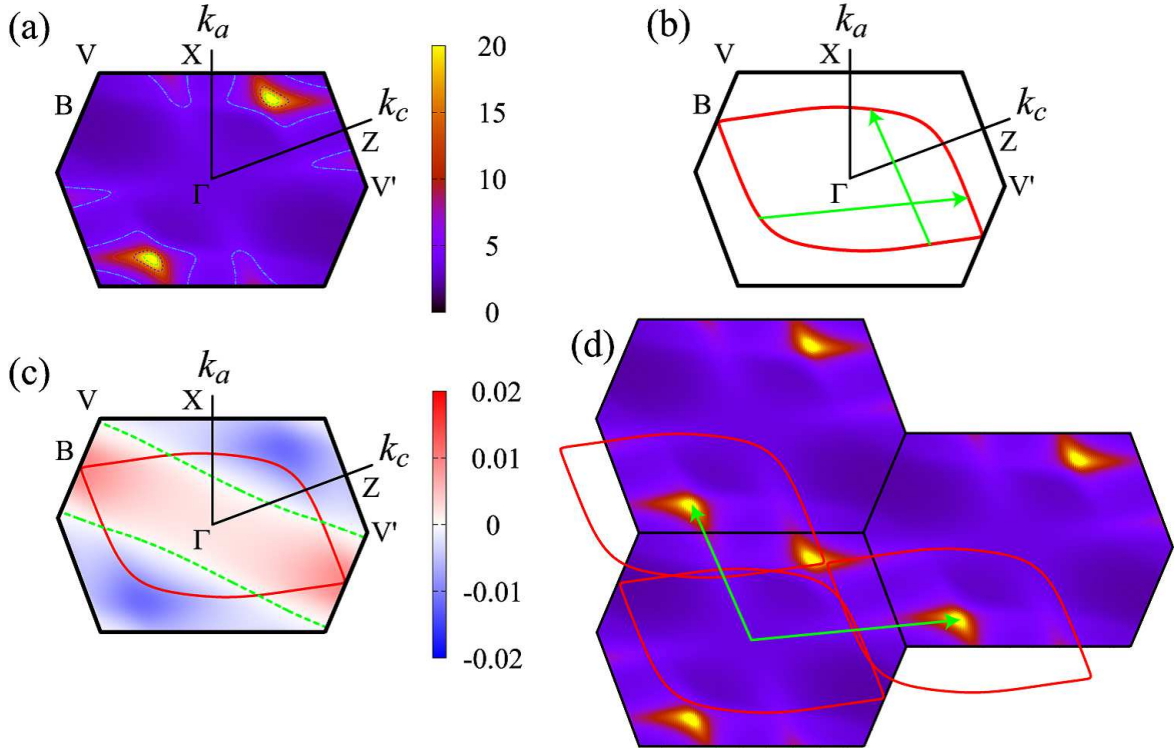}
 \caption{  
 (a) The calculated spin susceptibility. 
 (b) Arrows correspond to the 
peak position wave vector of the spin susceptibility 
(nesting vector of the Fermi surface).  
 (c) Spin singlet superconducting gap function 
 of the model for $\beta$-(BDA-TTP)$_{2}$SbF$_{6}$, 
 where $U=0.38$, $T=0.008$ in 128$\times$128. 
 For this parameter set, 
 the obtained $\lambda$ is 0.923. 
 (d) The relation between the nesting vector of the Fermi surface 
 and the spin susceptibility. 
 }
 \label{fig7}
\end{figure}

We now move on to the RPA analysis for the effective models.
To focus on the anion dependence, here we concentrate on the models  
obtained using the lattice parameters at room temperature.
We show in figure\ref{fig7}(a) the spin susceptibility 
in the Hubbard model of $\beta$-(BDA-TTP)$_{2}$SbF$_{6}$, 
where $U=0.38$[eV] and the temperature $T=0.008$[eV] here.
The wave vector at which the spin susceptibility 
is maximized is $ \textbf{\textit{k}}=\left( 0.45 \pi, 0.53 \pi \right)$. 
The arrows in figure\ref{fig7}(b) corresponds to the peak position of the 
spin susceptibility in (a), which can be considered as the nesting 
vector of the Fermi surface. 
In fact, there is a certain overlap between the original Fermi surface and 
the ones translated by these arrows, as shown in figure \ref{fig7}(d). 
The superconductivity is analysed by solving the 
linearized gap equation that takes into account the pairing interaction 
mediated by the spin fluctuations. 
The superconducting gap function of the spin-singlet pairing 
in figure \ref{fig7}(c) shows that the sign of the gap changes four times 
along the Fermi surface like in a $d$-wave gap. 
Considering that the $\Gamma$-B direction corresponds to 
the $a-c$ direction in the lattice structure, 
the obtained gap function indicates that 
the hot spots are present in the $k_{a} \pm k_{c}$ direction. 
The position of the nodes is different from those obtained in 
previous theoretical studies 
that adopt extended H$\ddot{\rm u}$ckel models 
(see the appendix) 
\cite{Nonoyama-Maekawa-JPSJ-77-094703,comment-Nonoyama-et-al,Suzuki-Onari-JPSJ-80-094704}, 
and hence is not consistent with the STM study
\cite{Nomura-Muraoka-PhysicaB-404-562}.  
On the other hand, the present result is indeed more 
consistent with the recent measurement of the anisotropy of $H_{\rm c2}$ 
in $\beta$-(BDA-TTP)$_{2}$SbF$_{6}$ 
\cite{Yasuzuka-Koga-ICSM2010-5Ax-10}. 
Further study is necessary to resolve 
the consistency between existing experiments and theories.

\begin{figure}[!htb]
 \centering
 \includegraphics[width=11.0cm]{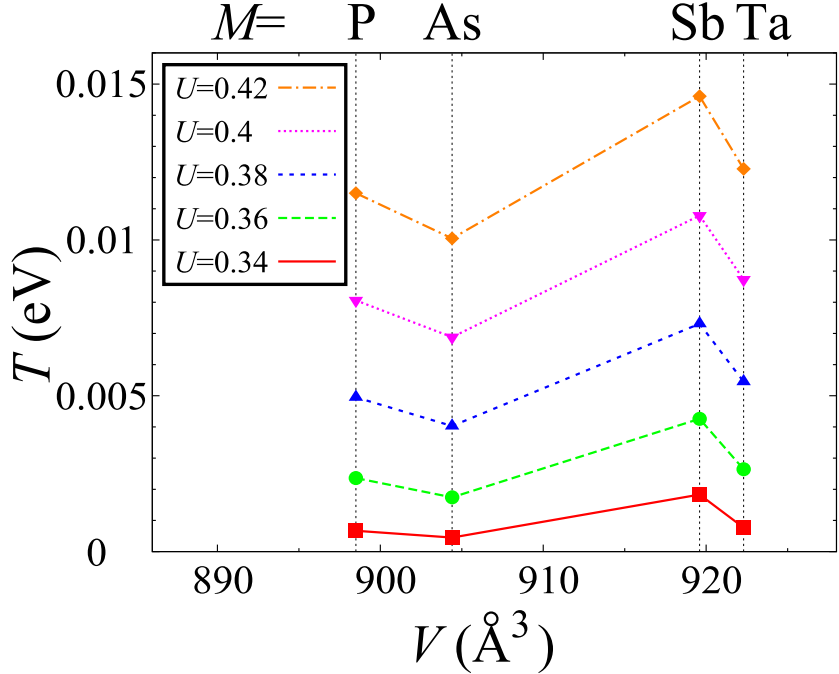}
 \caption{  
 $T_{\rm c}$ within RPA method 
 for various on-site interaction 
 as a function of the unit-cell volume for 
 $\beta$-(BDA-TTP)$_{2}M$F$_{6}$ ($M$=P, As, Sb, Ta) of ambient pressure. 
 }
 \label{fig8}
\end{figure}

$T_{\rm c}$ is obtained for the four anions for various on-site 
interaction $U$. 
figure \ref{fig8} shows $T_{\rm c}$ as functions of the unit-cell volume. 
There are several features found in the calculation results: 
(i) $T_c$ is the highest for Sb, 
(ii) $T_c$ somewhat decreases as we go from P to As, and 
(iii) $T_c$ goes down as we go from Sb to Ta, but it is still finite, close to that of P. 
These features are in accord with the anion dependence of 
DOS($E_{\rm F}$) or $W_{\rm SOMO}$ shown in Fig \ref{fig4}(c) or (b), 
but not $W_{\rm SH}$ or the unit-cell volume. 
Point (i) is indeed consistent with the experimental finding. 
Point (ii) can also give an understanding for the experimental 
observation that $T_c$ does not increase substantially as we go from P to As. 
This may at first sight be puzzling from the viewpoint of the 
increase in the unit-cell volume, 
from which one would expect an increase in the density of states. 
The key for the present understanding is the difference in the strength 
of the dimerization between P and As as mentioned previously. 
As for point (iii), $T_c$ of Ta being somewhat close to P, 
is not consistent with the experiments, 
where no superconductivity is observed for Ta. 
The discrepancy may be due to effects 
that are not taken into account in the present model (such as phonons), 
or in the RPA (more sophisticated electron correlation effects). 
On the other hand, the decrease of the $T_c$  
with the increase of the unit-cell volume found 
in the present calculation might partially be related to the loss of 
superconductivity.


\section{Conclusion}
\label{Conclusion}

In the present paper, we have obtained the 
\textit{ab-initio} band structure and 
the effective tight-binding model of 
$\beta$-(BDA-TTP)$_{2}M$F$_{6}$ ($M=$P, As, Sb, Ta), 
and studied the superconducting gap form as well as the anion 
dependence of $T_c$ assuming spin fluctuation 
mediated pairing.

The results of \textit{ab-initio} band calculation 
for $\beta$-(BDA-TTP)$_{2}$SbF$_{6}$ show that 
the direction of the major axis of the cross section of the Fermi surface 
lies in the $\Gamma$-B direction, in contrast to the 
extended H$\ddot{\rm u}$ckel result.
The density of states at the Fermi level 
is correlated with the band width of SOMO rather than 
SOMO+HOMO, and hence is not necessarily correlated with the 
unit-cell volume. 

To construct an effective tight-binding model 
that accurately reproduces the \textit{ab-initio} band structure, 
we need both nearest  and next nearest neighbor transfer energies. 
From the obtained transfer energies, 
the dimerization of $M=$As is smaller than that of $M=$P and Sb, 
and hence the SOMO band width of $M=$P is smaller than that of $M=$As 
despite the larger SOMO+HOMO band width.

The RPA calculation shows 
that the wave vector at which the spin susceptibility is maximized 
lies in the $a+c$ direction. 
The gap function of the superconductivity 
mediated by the spin fluctuations changes sign 
along the Fermi surface like in a $d$-wave gap, and the  
hot spots of the gap (gap maxima) is in the  $k_{a} \pm k_{c}$ direction, 
again different from the extended H$\ddot{\rm u}$ckel model. 
$T_{\rm c}$ of $M=$Sb is the highest among all the anions  
in qualitative agreement with experiments.

There are some remaining issues that should be resolved in future studies.
As mentioned above, the relation between the present Fermi surface, 
those obtained by extended H$\ddot{\rm u}$ckel calculation 
\cite{Yamada-Watanabe-JACS-123-4174,Yamada-Fujimoto-SM-153-373,
Nonoyama-Maekawa-JPSJ-77-094703,Suzuki-Onari-JPSJ-80-094704}, 
and those obtained in two experiments 
\cite{Choi-Jobilong-PRB-67-174511, Yasuzuka-Koga-JPSJ-short-notes} 
remains as an open question.
As for the band calculation, 
we consider that the difference between the present {\it ab-initio} result 
and the extended H$\ddot{\rm u}$ckel result is intimately related to 
the empirical parameter in the extended H$\ddot{\rm u}$ckel calculation 
that controls including/excluding the 3d-orbital 
of the sulfur atoms of the TTP skeleton donors 
\cite{Ouyang-Yakushi-JPSJ-67-3191,Kawamoto-Ashizawa-JPSJ-71-3059,
Drozdova-Yakushi-JSSC-168-497}. 
Actually, the extended H$\ddot{\rm u}$ckel result in which 
the 3d-orbital of the sulfur atoms is excluded 
\cite{Kawamoto-private-comm-3d} 
is similar to the present \textit{ab-initio} result. 
%

Although all experimental and theoretical studies suggest 
anisotropic superconducting gap, 
the form of the gap function is different between the 
present and previous theoretical studies based on extend H$\ddot{\rm u}$ckel band
\cite{Nonoyama-Maekawa-JPSJ-77-094703,Suzuki-Onari-JPSJ-80-094704}. 
Given the consistency between the recent 
experiment\cite{Yasuzuka-Koga-JPSJ-short-notes}  and 
the present calculation regarding 
the direction of the anisotropy of the Fermi surface, 
we believe that the gap maxima should lie in the $k_{a} \pm k_{c}$ direction 
{\it as far as the spin-fluctuation-mediated pairing mechanism is concerned}. 
If the nodes of the superconducting gap do not lie 
in this direction in the actual materials, 
then the present study would conversely suggest that 
a different pairing mechanism may be at work. 
In this sense, it is highly important to resolve the experimental controversy 
regarding the direction of the nodes in the superconducting gap
\cite{Nomura-Muraoka-PhysicaB-404-562, Yasuzuka-Koga-ICSM2010-5Ax-10}.

\ack
We acknowledge T. Kawamoto, M. Tsuchiizu, A. Kobayashi, Y. Suzumura, 
and K. Nomura for valuable discussions. 
This work is supported by Grant-in-Aid for Scientific Research from the
Ministry of Education, Culture, Sports, Science and Technology of
Japan, and from the Japan Society for the Promotion of Science.
Part of the calculation has been performed at the 
facilities of the Supercomputer Center, 
ISSP, University of Tokyo.

\appendix

\section{Spin susceptibility and pairing symmetry 
in the model based on the extended H$\ddot{\rm u}$ckel band}
\label{Appendix}

In this appendix, 
we show the calculation result obtained by applying the same RPA method to 
the Hubbard model based on the extended H$\ddot{\rm u}$ckel band. 
Although the present result is essentially the same as those obtained in 
previous studies 
\cite{Nonoyama-Maekawa-JPSJ-77-094703,comment-Nonoyama-et-al,Suzuki-Onari-JPSJ-80-094704} 
here we show the results in the same manner 
as in figures \ref{fig7} to make the comparison clearer.
Here, we concentrate on the  model for 
$\beta$-(BDA-TTP)$_{2}$SbF$_{6}$ \cite{Yamada-Watanabe-JACS-123-4174}. 
The parameter values are $U=0.34$[eV] and  $T=0.008$[eV].
\begin{figure}[!htb]
 \centering
 \includegraphics[width=15.5cm]{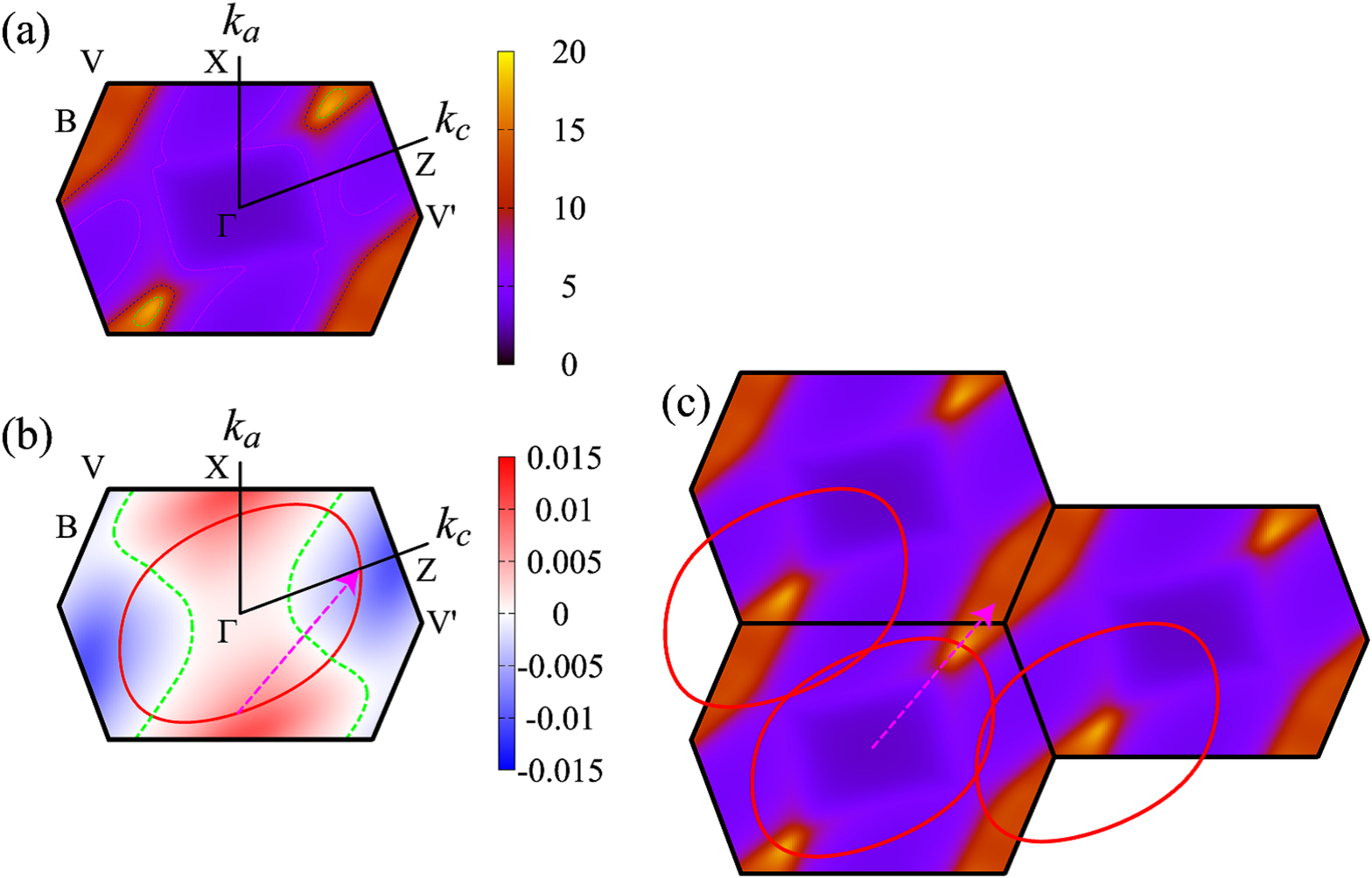}
 \caption{  
 (a) The calculated spin susceptibility. 
 (b) Spin singlet superconducting gap function 
 of the extended model H$\ddot{\rm u}$ckel 
for $\beta$-(BDA-TTP)$_{2}$SbF$_{6}$  
 with $U=0.34$, $T=0.008$ in 128$\times$128. 
 For this parameter set, 
 the obtained $\lambda$ is 0.997. 
 (c) The relation between the superconducting gap sign change 
 and the spin fluctuations. The dashed arrows connect the portions of 
 the Fermi surface that have the maximum gap with different signs.
 }
 \label{fig9}
\end{figure}

The calculated spin susceptibility is shown in figure \ref{fig9}(a). 
It can be seen that it has a broad maximum around the B point, 
indicating that there is no well defined nesting for this Fermi surface.
The spin-singlet superconducting gap function in figure \ref{fig9}(b) 
shows that the sign of the gap changes four times 
along the Fermi surface, 
but the direction of the nodes 
differs from that of the \textit{ab-initio} band model 
shown in figure \ref{fig7}(c). 
If we consider a wave vector that connects the two gap maxima positions 
on the Fermi surface with different signs of the gap, 
the vector falls on the broad maximum area in the 
spin susceptibility as shown in figure\ref{fig9}(c), 
indicating that the sign change is indeed due to the 
repulsive pairing interaction mediated by the spin fluctuations.
We note that the direction of the nodes obtained for this model is 
in the $k_{a} \pm k_{c}$ direction, which is in 
agreement with the STM measurement \cite{Nomura-Muraoka-PhysicaB-404-562}, 
but not with the in-plane anisotropy of $H_{\rm c2}$ 
\cite{Yasuzuka-Koga-ICSM2010-5Ax-10}.

\section*{References}


\end{document}